\documentclass[10pt,letterpaper]{article}
\usepackage{opex3,cite}

\usepackage{microtype}
\DisableLigatures{encoding = *, family = * }

\definecolor{forestgreen}{RGB}{34, 139, 34}

\newcommand{\Rb}{$^{85}$Rb }
\newcommand{\degree}{\ensuremath{^\circ}}

  \newcommand{\dd}{
      \mathop{}\mathopen{}\mathrm{d}   }

\begin{document}

\title{Temporally multiplexed storage of images in a Gradient Echo Memory}

\author{Quentin Glorieux*,  Jeremy B. Clark, Alberto M. Marino, Zhifan Zhou, and Paul D. Lett}

\address{Quantum Measurement Division, National Institute of Standards and Technology
and Joint Quantum Institute, NIST, and the University of Maryland, 
Gaithersburg, MD 20899, USA}

\email{*quentin.glorieux@nist.gov} 



\begin{abstract} 
We study the storage and retrieval of images in a hot atomic vapor using the gradient echo memory protocol.
We demonstrate that this technique allows for the storage of multiple spatial modes.
We study both spatial and temporal multiplexing by storing a sequence of two different images in the atomic vapor.
The effect of atomic diffusion on the spatial resolution is discussed and characterized experimentally.
For short storage time a normalized cross-correlation between a retrieved image and its input of 88~\% is reported.
\end{abstract}

\ocis{(270.0270) Quantum Optics; (270.5565) Quantum communications; (210.4680) Optical memories.}  


\section{Introduction}
Storage and retrieval of non-classical states of light has recently become a topic of interest due to the need for quantum memories (QM) in quantum network protocols \cite{Kimble:2008uv}.
Indeed, a QM is required for quantum repeaters which allow long distance distribution of entanglement \cite{Sangouard:2011bp,Sangouard:2008jc}.
In the perfect case, a QM should preserve the quantum state of a light pulse during the longest storage time possible.
A large number of techniques have been investigated (see \cite{Simon:2010wy,Lvovsky:2009vg} and references therein).
An attractive way to improve the performance of a QM is by multiplexing communication channels \cite{Munro:2010cn,Simon:2010cn}.
Temporal multiplexing can be used to increase the rate of entanglement distribution \cite{Afzelius:2009uc,Bonarota:2011tn}.
Similarly, a spatially multimode approach can, in principle, reduce the requirements on the storage time for a realistic implementation of a quantum repeater \cite{Lan:2009ts,Collins:2007cs}.
Experimentally, temporal and spatial multiplexing have been demonstrated independently with different techniques in atomic memories.
Temporal multiplexing, in an atomic medium, has been recently demonstrated using the gradient echo memory (GEM) technique \cite{Hosseini:2009fd}.
Correspondingly, using the electromagnetically induced transparency (EIT) technique, several groups have stored an image in an atomic vapor \cite{Surmacz:2008el,Shuker:2008bn,Vudyasetu:2008wv}.
In this paper, we demonstrate the ability of the GEM in an atomic vapor to store multi-spatial-mode images and we show that the resolution of the retrieved images is ultimately limited by the atomic diffusion of the atoms.

Given that there exists a straightforward method for the generation of multi-spatial-mode quantum states near the Rb atomic resonance \cite{boyer1,glx2}, we would like to eventually be able to store such states in an atomic vapor memory.  While a large number of other approaches to a quantum memory are available, our experiment is implemented using the promising technique of gradient echo memory in an atomic vapor introduced by Buchler and coworkers \cite{Hetet:2008wi}.
This technique has recently led to a recovery efficiency of 87~\% \cite{Hosseini:2011jv}, temporal multiplexing \cite{Hosseini:2009fd}, and storage of a weak coherent state with a recovery fidelity above the no-cloning limit \cite{Hosseini:2011iv}.  Thus, the
GEM is a promising candidate for quantum memory because it has a very high recovery efficiency not achieved yet with other techniques in a hot atomic vapor.  It is also relatively easy to implement, not requiring laser-cooled atoms or cryogenics.
The ability of the GEM to store and recover multi-spatial-mode images, and to do so with high spatial fidelity has, however, not yet been demonstrated, and the present experiments are designed to verify the image storage properties of the GEM in much the same way as has been done for other techniques.

It is also interesting to combine multiplexing techniques for both time and space.
So far storing multiple images have been demonstrated only in solid-state systems.
Several experiments in holographic memories have shown to be able to record trains of images by storing the information in the atomic population \cite{Sasaki1995273,Mitsunaga:94} and have opened the way to photon-echo protocols.
Here, we report  the coherent storage and retrieval of a sequence of two different images using an atomic vapor gradient echo memory.

\section{Experimental setup}
\begin{figure} []
		\centering
			\includegraphics[width=\textwidth]{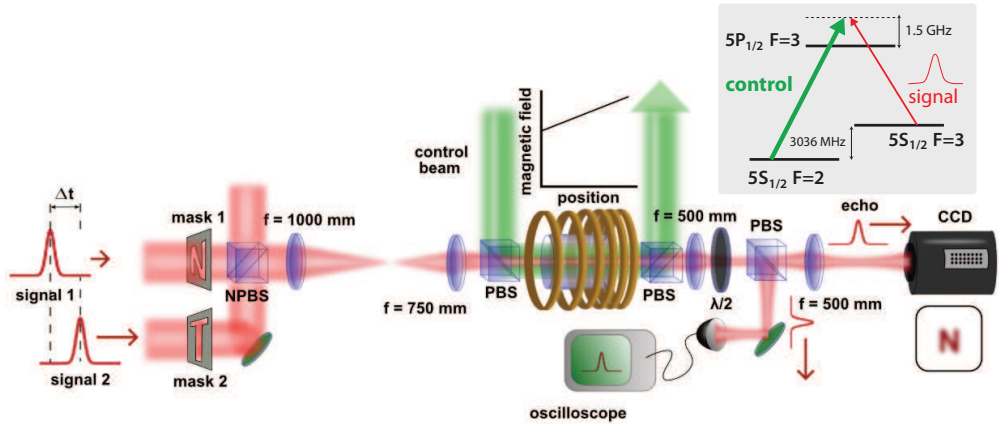} 
		 \label{fig1}
		\caption{Schematic of the experimental setup for image storage in a GEM.
		The signal beam is previously shaped temporally with an acousto-optic-modulator (not shown).
		The spatial profile is fixed using a mask in the beam path and the mask is then imaged into the \Rb cell with a magnification of 0.75.
		A control field is overlapped with the signal using a polarizing beam splitter (PBS), and finally, the retrieved image is recorded using a fast--gated intensified camera.
		A small fraction of the signal is sent to a fast photodiode for monitoring the spatially-integrated retrieved power.
		For the storage of two consecutive images, the two signal beams are combined with the same polarization on a non-polarizing beam splitter. 
		Inset: three-level system for $^{85}$Rb.}	
\end{figure}

In our GEM experiments a light pulse is sent into a \Rb atomic vapor which can be considered, for simplicity, as a collection of three-level systems in a $\Lambda$ configuration.
A spatially-dependent Zeeman shift is applied using a linearly varying magnetic field along the axis of propagation.
For a total Zeeman shift larger than the pulse frequency bandwidth, a high optical density, and the presence of an intense co-propagating control field that drives the Raman transition, each frequency component of the pulse can be absorbed.
The photonic excitation is therefore transferred into the long lived collective coherence of the hyperfine atomic ground states.
Similar to nuclear magnetic resonance or photon echo techniques, the spectral components of the signal are mapped into the medium along the length of the cell.
After the excitation, the collective dipole dephases due to the inhomogenous magnetic field.
It is possible, however, to rephase the collective excitation by reversing the magnetic field gradient.
The time evolution is then reversed, and when all the dipoles have rephased the retrieved light pulse emerges in the forward direction provided that the control field is again present.
As demonstrated in \cite{Hosseini:2011jv}, this technique preserves a phase relationship between the input and retrieved pulses and is therefore a coherent storage.

In our experiment, the transverse profile of the signal field is shaped using a mask and imaged into the atomic medium.
For a control field much wider than the signal, the transverse profile of the collective excitation of the atomic coherence directly mimics the profile of the signal field, so spatial information can be stored in the atomic memory.
In this paper we first describe the experimental setup and present results on the storage of two images into the atomic medium.
We then introduce the criterion of similarity, based on a normalized cross--correlation, to analyze the spatial fidelity of the retrieved images with respect to its input reference image.
The crosstalk between the first and second retrieved images is investigated with regard to this criteria.
Finally, we use a resolution chart to measure and quantify the effect of atomic diffusion at a given buffer gas partial pressure on the storage of spatial information.  We investigate primarily the transverse spatial diffusion, as longitudinal diffusion along the direction of the field gradient effectively prevents atoms from contributing to the coherent echo signal, and we show that the resolution of the retrieved images primarily depend on the storage duration, as expected and similar to what is found with EIT techniques.


The experimental setup is shown in Fig.~1.
The light from a Ti:sapphire laser, blue detuned by 1.5~GHz from the 5S$_{1/2}$,~F=2~$\rightarrow$~5P$_{1/2}$,~F=3 transition, is used as a control field for the Raman coupling.
The signal beam is generated using a double-passed 1.5~GHz acousto-optic modulator to downshift the frequency by the amount of the hyperfine splitting (3.036~GHz).
The signal and control beams are overlapped with crossed linear polarizations in the 5~cm long memory cell.
The cell contains isotopically pure \Rb and 667~Pa (5~Torr) of Ne buffer gas and is heated up to 80~\degree C.
A bias magnetic field of 1~mT (10~G) is applied to the cell to split the ground state Zeeman sublevels and select a specific three-level system (see Fig. 1 inset).
The Raman absorption line is broadened by a 15~$\mu$T/cm magnetic field gradient we apply in the propagation direction.
The spatial profile of the signal beam is shaped by placing a mask in its path and imaging it into the memory cell.
In order to store two different images, two distinct signal beams have to be shaped independently.
The temporal profile of each signal pulse is Gaussian with a full width at $1/e^2$ of 1.1~$\mu$s.
The storage duration is set by the delay between the input pulse and the flip of the magnetic gradient.
The flip duration is less than 1 $\mu$s.
After the cell, the control field is filtered with a polarizing beam splitter.
A fast--gated intensified camera records the time--integrated light intensity during 100~ns wide frames.
For convenience, a small fraction of the retrieved image beam is sent to a photodiode for recording the retrieved power as function of time.


\begin{figure}	[]	
		\centering
		\includegraphics[width=0.70\columnwidth]{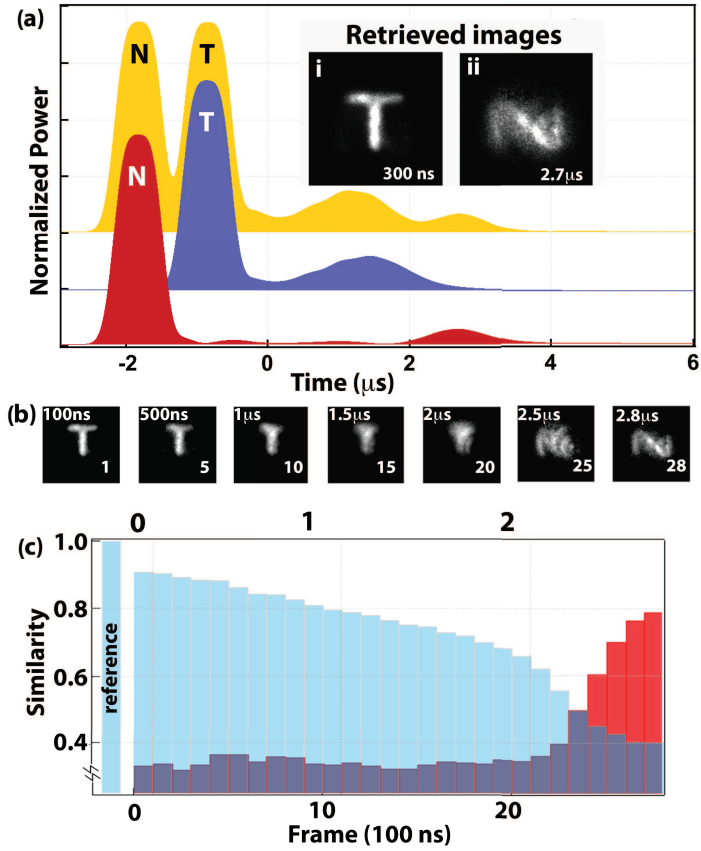} 
		 \label{fig3}
		\caption{Storage and retrieval of two images in a GEM. 
		(a) The spatially-integrated intensity from the retrieval of the single letter T (blue), and N (red).
		The input pulse is at negative time and the retrieved pulse is at the symmetric positive time after the magnetic field gradient flip.
		The yellow curve shows the storage of two images, and the curves are vertically displaced for clarity.
		Retrieved images at time 0.3~$\mu$s and 2.7~$\mu$s are presented, respectively, in Fig.~2(a), i and ii.
		(b) The detailed timeline of the retrieved images for 7 frames. The frame number is indicated in the images and the time corresponds to the time after the magnetic flip.
		(c) $S_N$ (red) and $S_T$ (blue) as function of the frame number.
		The reference for N or T has a similarity of 1 and is plotted before 0. }	
\end{figure}

\section{Temporal and spatial multiplexing}
In Fig.~2(a) we present the timeline of the storage and retrieval of two consecutive images.
The amplitude of the retrieved pulse is normalized to the input amplitude.
In this particular configuration (control beam diameter of 3~mm and power 120~mW, and the probe beam diameter of 1~mm) we report a retrieval efficiency of 8~\%.
Improvements to this can be obtained by using a longer memory cell, thereby improving the optical depth for the Raman absorption. The optical depth on resonance ~ 200, nevertheless, the probe beam is far detuned from resonance, and the absorption is due to a 2-photon Raman absorption.
In the results presented here, the absorption of the input beam is on the order of 30~\%, leading to a maximum theoretical retrieval efficiency of 9~\%. (We have measured a recovery efficiency of up to
62~\% for a 6.5ms storage time in a similar configuration by using a 20 cm long memory cell.)
As we are interested in the spatial properties of the GEM in this work we do not focus on the overall efficiency of the process.
We first check that a single image can be stored, and optimize the alignment independently for the two different signal beams corresponding to the images of the letters N and T (the red and blue traces in Fig.~2(a), respectively).
To store both images in the atomic memory, the two input pulses are combined on a beam splitter with a delay of 1~$\mu$s between them.
Two frames from the gated intensified camera of the retrieved images after a storage time of 0.5~$\mu$s (frame 3) and 4.5~$\mu$s (frame 27) are presented, respectively, in Fig.~2(a), i and ii.
As GEM in this configuration is a first-in-last-out memory \cite{Hosseini:2009fd}, the first retrieved image corresponds to the letter T and the second one to the letter N.

A more detailed timeline of 7 frames for the retrieved images is included in Fig.~2(b).
This figure shows the intensity profile of the retrieved light for different times and clearly exhibits an overlap between the two letters in frame 25.
To study the overlap quantitatively we define the similarity $S$ to be the cross--correlation of a given frame with a reference image, normalized by the square root of the auto--correlation product :
\begin{equation}
S=\frac{\sum_{i,j}N_{ij}^{in}N_{ij}^{echo}}{\sqrt{\sum_{i,j}{(N_{ij}^{in})^{2}}\sum_{i,j}{(N_{ij}^{echo})^{2}}}},
\end{equation} 
where $N_{i,j}^{in}$ is the intensity recorded for pixel {i,j} of the input pulse image and $N_{i,j}^{echo}$ for the retrieved image.
The reference image is selected from among the frames of the input pulse for each letter.
In  Fig.~2(c) we plot the similarities for the retrieved images with respect to the N and the T reference images (respectively denoted $S_N$ and $S_T$).
For reference, the similarity between the T and the N input images is 35~\%. 
The evolution as function of time is given for 28 non-overlapping frames representing successive 100~ns time intervals.
The first frame is taken 100~ns after the magnetic field gradient is switched.
We expand the retrieved pulse compared to the input (expansion ratio of 1.4) using a different magnetic field gradient for the recovery\cite{Hosseini:2009fd}.  This allowed for a slightly longer diffusion time and made it easier to measure in the retrieved images.
The total storage time is therefore given by twice the time measured from the magnetic field flip corrected by the expansion ratio.
For the retrieved image the initial value of $S_T$ is 88~\% and it decreases at a rate of 1.1~\% per frame due to the atomic diffusion that blurs the image.
After 19 frames the value of $S_T$ starts to drop faster, at a rate of 4.6~\% per frame.
On the other hand, $S_N$ is initially low (35~\%) and after 21 frames it suddenly rises at a rate of 8.5~\% per frame.
After frame 28 $S_N$ reaches 78~\%, corresponding to when the second echo is retrieved from the memory.
The two curves cross during the 24th frame with $S_T$=47~\% and $S_N$=51~\%.
This is close to a value of 50~\%, which we take to be a threshold for distinguishing the images.
As shown by the yellow curve of Fig.~2(a), there is a temporal overlap between the two input images.
The similarities corresponding to this overlapping input frame are $S_T$=55~\% and $S_N$=48~\%.
On the other hand, the frame just before can be identified as an N and the frame just after as a T, with a difference $D=|S_N-S_T|>0.5$.
The retrieved images mimic this behavior with an overlap during frame~24.
The value of $D$ immediately before (frame 23) and after (frame 25) is greater than 0.15.
These similar qualitative behaviors suggest that there is no significant crosstalk between the two images and that the reduction of $D$ between the input and the output is a consequence of the blurring of the images.

\section{Effect of  atomic diffusion}

\begin{figure}	[]	
		\centering
		\includegraphics[width=0.70\columnwidth]{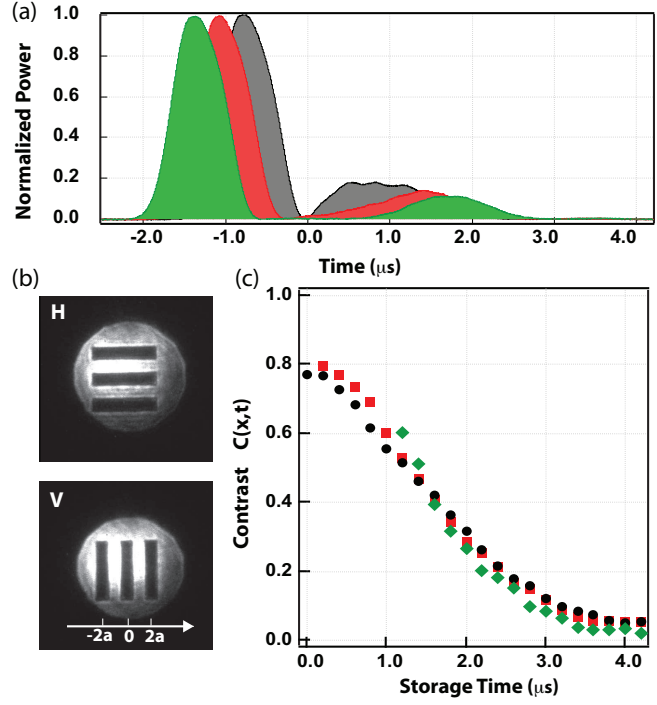}
			\caption{(a) The input and retrieved pulses for three different delays.
		 	Time t=0 is taken to be the time of the magnetic field flip.
			The maximum of the input pulses are at -1.4~$\mu$s, -1.1~$\mu$s, -0.8~$\mu$s (respectively green, red, and black curves).
			(b) The input images for horizontal (H) and vertical (V) lines. 
			$1/2a$ is the spatial frequency of the bars, where the distance between a black and a white line is denoted $a$.
			(c) Contrast for the retrieved image (vertical) as function of storage time for 21 frames. 
			The symbol colors correspond to those in a).}	
\end{figure}

It has been proposed that the decay of $S$ during the storage in an EIT based memory is mainly due to atomic diffusion \cite{Shuker:2008bn,Firstenberg:2008fx}.
We show that our experiment using a GEM technique is consistent with this statement and demonstrate that the expected spatial resolution of the retrieved images can be predicted from the buffer gas partial pressure and the storage time. 
The internal dynamics of the atoms are described by the optical Bloch equations \cite{cohen}.
In the presence of a buffer gas, the Rb atoms are subject to velocity changing collisions with the gas that are assumed to preserve the internal state.
The atomic excitation $\rho(x,y,z,t)$ therefore diffuses in the transverse directions as follows :
\begin{equation}\label{diffu}
\frac{\partial}{\partial t}\rho(x,y,z,t)=D\nabla^2 \rho(x,y,z,t),
\end{equation} 
where D is the diffusion coefficient.
The diffusion along the propagation axis is neglected as it induces only loss in the process and does not affect the spatial information in the transverse plane. 
Indeed, if one atom diffuses along the direction of the magnetic gradient, it will not rephase at the same time as others and therefore will not contribute to the expected retrieved image.
Using the Green's function formalism, the evolution in time and space of a single point initially at the position $X'=(x',y')$ can be calculated.
Thus the evolution of the atomic excitation from an arbitrary initial condition can be derived by knowing the spatial distribution $\rho(x',y',0) $ of the excitation at t=0  as follows : 
\begin{equation}\label{soldifu}
\rho(x,y,t)=\frac{1}{\sqrt{4\pi D t}} \int  e^{\frac{-(x-x')^2-(y-y')^2}{4Dt}}\rho(x',y',0) \dd x' \dd y'.
\end{equation} 
To evaluate quantitatively the effect of diffusion on the spatial resolution we stored and retrieved images of a resolution test chart consisting of a group of three bars oriented vertically or horizontally as shown in Fig.~3(b). 
The resolution chart is imaged into the memory with a magnification of 0.75.
For three different pulse delays relative to the magnetic field switching time (Fig.~3(a)) we record 100~ns frames during the retrieval process.
The contrast $C(t)$ is defined as:
\begin{equation}\label{soldifu}
C(t)=\frac{I(x=a,t)-I(x=0,t)}{I(x=a,t)+I(x=0,t)},
\end{equation} 
where $x$ can refer to the vertical or the horizontal direction and $a$ is the width of a stored line taking into account the magnification (see Fig.~3(b)).
At $t=0$, $x=0$ refers to the center dark line and $x=a$ to the center of a bright line.
$I(x,t)$ is the intensity recorded by the camera at position $x$ for a frame recorded at time $t$, integrated over the extent of the pattern in $y$.
With this definition $C$ can fall below zero for a long storage time as the atoms diffuse from bright lines to dark lines resulting in $I(x=0) > I(x=a)$.
As shown in Fig.~3(c), the contrast for a given spatial frequency does not depend on the time between the input pulse maximum and the magnetic field switching, but only on the total storage time.
We have also independently checked that  the orientation of the lines does not affect the contrast.
The fact that the only parameter that affects the contrast is the storage duration strongly suggests that diffusion is the main source of the degradation of the resolution.

\begin{figure}	[]	
		\centering
			\includegraphics[width=0.75\columnwidth]{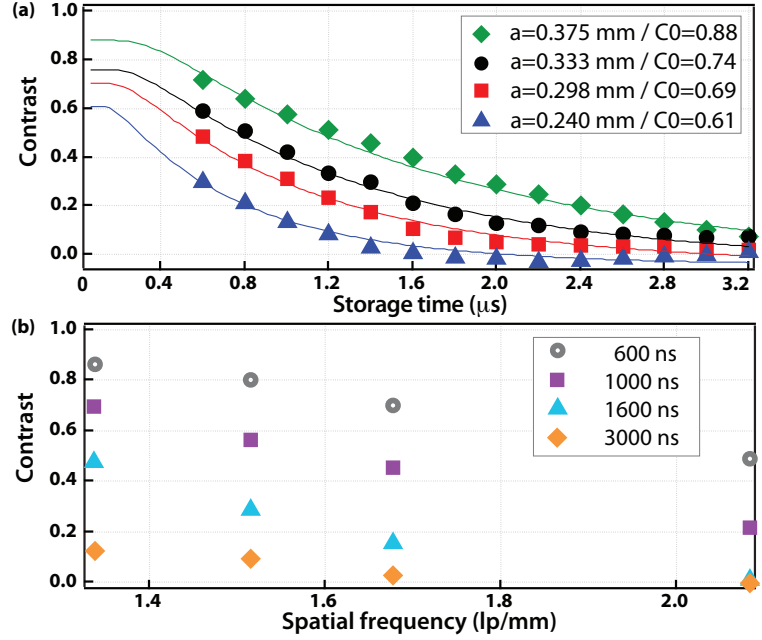} 

		 \label{fig13}
		\caption{(a) Contrast and theoretical fit for four images such as shown in Fig.~3(b), with bar widths, $a$, as indicated in the figure.
		The fits are done without free parameters using Eq~(3). D=105  cm$^2$/s, $t_0$=0 and  $C_0$ is given in the legend. 
		(b) Modulation transfer function for four different storage times. }	
\end{figure}

To confirm this observation we have compared 4~masks with different spatial frequencies.
The resolution of the mask varies from of 1 to 1.6 line pairs/mm, resulting in lines of 375~$\mu$m to 240~$\mu$m in the image plane in the memory after taking into account the magnification.
In Fig.~4a, we plot the contrast as function of storage time for the different masks and the fit of the data to the model of Eq.~(3).
In order to take into account independently the quality of our optical system and the contrast degradation due to the memory we have measured the initial contrast of the input pulse $C_0$.
$C_0$ is measured for the different masks and is a fixed normalization parameter for the fit.
We take the value of the diffusion coefficient for 667~Pa of Ne buffer gas at 80~\degree C to be 105~cm$^2$/s as previously reported \cite{Franz:1965vf,Franzen:1959wj}, and therefore no free parameters are allowed for the fit.
The good agreement between our theoretical model based on atomic diffusion and the experimental data confirms the key role of diffusion in the contrast of the retrieved image for the GEM technique.
In Fig.~4(b) we plot the modulation transfer function (contrast as a function of spatial frequency) for different storage times.
This graph  could be a useful benchmark for spatial channel multiplexing for a quantum communication network, as the three lines of the test chart can be seen as three parallel information channels.
Since the contrast needed for a given multi-spatial-mode quantum repeater protocol can be defined a priori,  Fig.~4(b) provides the maximum spatial frequency allowed for  a fixed storage time and therefore the number of spatial channels which can be simultaneously used to store quantum information.

\section{Conclusion}
We have shown that a multi-spatial-mode image can be stored using the GEM protocol.
Moreover, we have demonstrated that multiple images can be stored and retrieved at different times, allowing the storage of a short movie in an atomic memory.
This opens the way to multiplexing simultaneously in time and in space for future quantum memory applications.
We confirm that the main limitation of this technique, similar to that in EIT-based memory \cite{Shuker:2008bn,Firstenberg:2008fx}, is the atomic diffusion during the storage time.
We expect that this could be overcome by using a cold atomic sample \cite{Lukin:2003tca,Fleischhauer:2002wia} or mitigated by storing the Fourier transform of the image into the memory as suggested in  \cite{Shuker:2008bn,Zhao:2008iz}.
Finally, as the different spectral components of the input signal are mapped along the length of the cell with GEM, unlike EIT, it would be interesting to investigate if there is an effect of longitudinal diffusion on the signal spectral properties and noise.

This work was supported by the AFOSR.
Q.G. and P.D.L. thank Ping Koy Lam and Ben Buchler of the Quantum Optics Group at ANU for hosting a visit by Q.G.
Q.G. thanks all the members of this group for fruitful discussions.

\end{document}